\documentclass[scmsmall]{acmart}
\AtBeginDocument{%
  \providecommand\BibTeX{{%
    \normalfont B\kern-0.5em{\scshape i\kern-0.25em b}\kern-0.8em\TeX}}}

\copyrightyear{2025}
\acmYear{2025}
\setcopyright{acmlicensed}\acmConference[CHI '25]{CHI Conference on Human Factors in Computing Systems}{April 26-May 1, 2025}{Yokohama, Japan}
\acmBooktitle{CHI Conference on Human Factors in Computing Systems (CHI '25), April 26-May 1, 2025, Yokohama, Japan}
\acmDOI{10.1145/3706598.3714252}
\acmISBN{979-8-4007-1394-1/25/04}

\usepackage{subfiles}
\usepackage[normalem]{ulem}
\usepackage{graphicx}
\usepackage[skins]{tcolorbox}
\usepackage{enumitem}
\usepackage{multicol}
\usepackage{colortbl}
\usepackage{xcolor}
\emergencystretch=\maxdimen
\definecolor{Gray5}{gray}{0.95}
\definecolor{green}{HTML}{D9EAD3}

\newcommand{\edit}[1]{{\textcolor{black}{#1}}}

\begin{document}

\title[Epistemic Autonomy]{\edit{Moving Towards Epistemic Autonomy: A Paradigm Shift for Centering Participant Knowledge}}

\author{Leah Hope Ajmani}
\affiliation{%
  \institution{University of Minnesota}
  \city{Minneapolis}
  \state{Minnesota}
  \country{USA}}
\email{ajman004@umn.edu}
\author{Talia Bhatt}
\affiliation{%
  \institution{Trans/Rad/Fem}
  \city{Mumbai}
  \country{India}}
\email{taliabhattwrites@gmail.com}
\author{Michael Ann Devito}
\affiliation{%
  \institution{Northeastern University}
  \city{Boston}
  \country{USA}}
\email{m.devito@northeastern.edu}

\renewcommand{\shortauthors}{Leah Hope Ajmani, et al.}

\begin{abstract}
Justice, epistemology, and marginalization are rich areas of study in HCI. And yet, we repeatedly find platforms and algorithms that push communities further into the margins. In this paper, we propose epistemic autonomy--—one’s ability to govern knowledge about themselves---as a necessary HCI paradigm for working with marginalized communities. We establish epistemic autonomy by applying the transfeminine principle of autonomy to the problem of epistemic injustice. To articulate the harm of violating one’s epistemic autonomy, we present six stories from two trans women:  (1) a transfem online administrator and (2) a transfem researcher. We then synthesize our definition of epistemic autonomy in research into a research paradigm. Finally, we present two variants of common HCI methods, autoethnography and asynchronous remote communities, that stem from these beliefs. We discuss how CHI is uniquely situated to champion this paradigm and, thereby, the epistemic autonomy of our research participants.
\end{abstract}



\received{September 2024}
\received[revised]{Decemeber 2024}
\received[accepted]{January 2025}

\maketitle

\section{Introduction}
As \edit{Human-Computer Interaction (HCI)} researchers have increasingly testified about their experiences of being transgender, Black, Indigenous, and disabled, there is a clear theme of fighting against pervasive erasure and disempowerment within computing systems (e.g., ~\cite{Haimson2020-aq, Erete2021-ul, Lazem2022-ft, Ymous2020-uc}). In many cases, both researchers and the users of our systems have expressed a sense of powerlessness. \edit{The fundamental changes these communities call for seem more like a distant future rather than a close reality.} This feeling of powerlessness serves a structural function; it is crucial to the status quo of patriarchy, racism, and ableism that marginalized communities internalize this message of disempowerment. Therefore, it is essential that HCI, as Computer Science's most human-facing subdiscipline, foster precisely the opposite.

Creating power here is the process of creating \textit{knowledge.} There is power in knowing oneself and being known by others on one's own terms ~\cite{Fricker2007-zh, Jackson2018-lo, Hogeland2016-gj}. However, the restrictive influences of whiteness, patriarchy, and colonialism are deeply embedded in our current knowledge processes. \edit{These theories of knowledge} tend to favor a search for one objective reality, provable only through means those already in power deem acceptable~\cite{Cunningham2022-zv, Kannabiran2023-qd, Rorty1985-su}. This dynamic results in a world where it is nearly impossible for those in the margins to know and assert themselves on their terms. Furthermore, we, as researchers, inadvertently spread a message that those in the margins are worthy of being observed but not worthy of being observers. In doing so, we further a prejudiced ideology: \textit{epistemic injustice}.

Epistemic injustice is the stripping of one’s rights to be a knower, such as the right to be an authority on one’s experience~\cite{Fricker2007-zh}. Sociotechnical systems are ripe venues for epistemic injustice, as they often have their own rules, norms, and politics~\cite{Reddy2023-zy}. Epistemic injustice rapidly compounds for those with intersecting marginalized identities. One key example of this phenomenon is how sociotechnical systems act on transgender women. Trans women are subject to egregious forms of epistemic injustice as they are repeatedly questioned as authorities of their own gender identity and health information. While online spaces can provide a venue for sharing information and building community, they have also formed algorithmic traps that actively endanger trans women seeking to use said systems~\cite{DeVito2022-ee}. More broadly, trans health and cultural information often get unduly flagged as misinformation~\cite{hamison_disproportionate_2021}. These dynamics are concerning given that a systemic lack of online health knowledge contributes to disproportionately high suicide rates in trans populations~\cite{Moody2015-ry, McNeil2017-by, Tebbe2016-lm}. At its worst, epistemic injustice is a violent act---propagating the gatekeeping and discrediting of life-altering knowledge for trans women. As the amount of information, curation, and subsequent moderation on the internet grows, so does the embeddedness of existing power structures around knowledge, with dire practical consequences.

Given that HCI research deeply informs the design of these platforms, we risk propagating epistemic injustice through our practices. At the same time, HCI is uniquely positioned to fight this same injustice, as work on how research interacts with justice, epistemology, and identity has a rich history in our field. \edit{In the past decade, we have seen a} new wave of HCI research intended to interrogate power structures, evaluate our research practices, and---ultimately---build more just technologies~\cite{Bardzell2011-ln, Sondergaard2023-lg, Menking2021-ic, O-Neill2024-ra}. And yet, time and time again, we find platforms and algorithms that push communities further into the margins~\cite{walker_more_2020, DeVito2018-gr}. This phenomenon is unsurprising, given how transphobia, misogyny, and patriarchy are deeply woven into our social fabric. However, recent work suggests that sociotechnical systems can serve as sites of resistance, a way to rip the stitches and weave a new fabric~\cite{Ajmani2024-ll}. Furthermore, sociotechnical research itself can further this mission. As research is considered ``canon knowledge,'' we all have the opportunity to push against epistemic injustice through our beliefs and practices. To that end, we ask: \textit{what does an orientation of HCI research towards fighting epistemic injustice look like?}

In this paper, we propose a re-orientation of HCI towards prioritizing the \textbf{epistemic autonomy} of our research subjects and their communities. \edit{We define epistemic autonomy as one’s ability to govern how they know themselves and how they are known to the world. We specifically build on \textit{autonomy} as articulated in transfeminism~\cite{Koyama2020-zi}---drawing heavily on the experiences from the two transfem authors of this paper.} After establishing what epistemic autonomy means in research settings, we demonstrate the serious consequences of violating one's autonomy. Using storytelling methods~\cite{Ogbonnaya-Ogburu2020-wt}, we include personal stories from (1) an online community context and (2) an academic context within HCI. Inspired by previous work on making CHI more inclusive and progressive~\cite{Bardzell2011-ln, Ogbonnaya-Ogburu2020-wt, Frauenberger2020-ol}, we frame epistemic autonomy as a research paradigm. While we name an epistemic problem here (i.e., epistemic injustice), we believe the solution also requires a fundamental inquiry into our ontological and methodological beliefs. Finally, we apply our paradigm to two current methods in HCI, autoethnography and asynchronous remote communities (ARCs). We outline variants of these methods, scaffolded autoethnography and member-checked ARCs, to demonstrate how researchers can better prioritize epistemic autonomy in their current research practices. In sum, this paper contributes a \textbf{theory of epistemic autonomy}, a \textbf{critique of current HCI} research beliefs, and two \textbf{methods for fighting epistemic injustice} within our research.

\section{Related Work}
\subsection{Epistemic Injustice} 
Concisely, \textbf{epistemic injustice} is the systemic stripping of one's due right to be a knower. \edit{Popularized by Miranda Fricker~\cite{Fricker2006-jm}, epistemic injustice is a synthesis of feminist work on how knowledge-oriented processes are value-laden (e.g., \citet{Haraway1988-ch}'s theory of situated knowledge, \citet{Harding2001-sj}'s standpoint epistemology, and \citet{suchman1993categories}'s politics of categories)}. Epistemic injustice comes in two forms: (1) testimonial injustice and (2) hermeneutical injustice. 

Testimonial injustice occurs when a speaker is not given due credibility because of prejudices against them. For example, healthcare sites continuously discount women's testimonies about their symptoms, pain levels, and bodily experiences~\cite{Carel2014-ep, Kidd2017-dl, Cohen_Shabot2021-sn}. Notably, trans healthcare practices exacerbate this injustice, where the stakes are often life-changing gender-affirming care~\cite{kcomt_profound_2019, drabish_health_2022}. In HCI, previous work has found that online spaces propagate and amplify the systemic silencing of transfem testimony~\cite{Ajmani2024-ll}, while social media platforms disproportionately remove trans health and wellbeing information overall due to campaigns of false reporting and harassment~\cite{hamison_disproportionate_2021}. The result is a reluctant exodus of transfems from these unsafe and unjust online spaces~\cite{DeVito2022-ee}. In this way, the systemic silencing of transfem testimony wipes the internet clean of crucial health information and erases transfems as knowers. 

Hermeneutical injustice occurs when one is stripped of the resources to interpret one's own experiences. Hermeneutical resources are often terms and concepts that capture a shared experience. For example, the concept of ``sexual assault'' was a pivotal hermeneutical resource for survivors because they could now describe, interpret, and reckon with the violent acts they endured~\cite{Fricker2007-zh}. Online communities are a unique venue for creating shared vocabularies and concepts. For example, the Wikipedia community has a rich ecosystem of terms that guide constructive deliberation~\cite{Geiger2010-ny}. On TikTok, users have created a new vocabulary of ``algospeak'' to discuss mental health topics without triggering the automoderator~\cite{Klug2023-qw}. While communities are creating their own set of hermeneutical resources, research has its own vocabulary rooted in positivist traditions. In this paper, we focus on the injustice that occurs when we---as researchers---do not adequately consider, validate, and empower the hermeneutical resources \textit{of the community}.

\subsection{Autonomy and Transfeminism}
\textbf{Autonomy} broadly describes one's ability to self-govern. For example, an autonomous vehicle is a machine that can self-regulate to move through the world without causing accidents. \edit{In this paper, we use an intersection of autonomy as articulated in HCI and autonomy as articulated in transfeminist philosophy.} In HCI,~\citet{Friedman1996-kx}'s work on Value Sensitive Design (VSD) initially listed autonomy as one of the key values to consider within technology design~\cite{Friedman2013-ge}. \edit{In VSD, autonomy is both a motivator for choosing certain research approaches---such as participatory methods---and an implication of research outcomes.} Recent work in HCI articulates autonomy as a precursor to other design ideals, such as algorithmic literacy. For example,~\citet{DeVito2021-td} centered autonomy as a goal, holding that for a person to be algorithmically literate, they must be able to pursue their own goals in their own ways.

Transfeminism builds on the fundamental feminist call for equity. \edit{In transfeminism, providing autonomy is a way of providing tools to defend equity. For example, affording trans women the simple ability to autonomously assert their gender (the core "trans women are women" ~\cite{Serano2007-yu}).} In fact, early transfeminist writing focused on confronting \textit{``social and political institutions that inhibit or narrow our individual choices''}~\cite{Koyama2020-zi}. While the transfeminist articulation of autonomy mainly centers on bodily autonomy, it is clear from the work above that the call is about fundamental and institutional change. Here, autonomy is a foundational principle of empowerment.

\edit{In this paper, we use the robust articulation of autonomy from transfeminist literature---one's ability to govern the fate of one's own life without coercion or violence~\cite{Hines2019-fj, Koyama2020-zi}. It is important to recognize how this articulation came about. Transfeminists have had to center and double down on autonomy because of how much of the world acts as a direct threat to transgender women's autonomy. The stigma around transness and gender queerness has stripped trans folks of their ability to define who they are~\cite{White-Hughto2015-cu}. The rampant silencing of medical information and trans experiences online has taken away trans folk's ability to govern their care~\cite{Ajmani2024-ll}. Previous work has found a correlation between these dynamics and the disproportionately high suicide rates of trans people~\cite{McNeil2017-by}. Issues of autonomy are life and death for trans folks, particularly trans women.}

\edit{Though autonomy centers on individual choice, transfeminism holds that autonomy is also relational~\cite{Bettcher2018-os}. Further articulated in Black feminism and disability justice scholarship, relational autonomy focuses on community care and autonomy as coupled concepts~\cite{Christman2004-fw, Davy2019-xy, Mackenzie2014-px}. For example, increasing bodily autonomy is part of the larger community-oriented transfeminist movement~\cite{Koyama2020-zi}. In this paper, we highlight both individual forms of epistemic autonomy, such as an individual's capacity to be a knower, and relational forms of epistemic autonomy, such as a community's capacity to create their own vocabulary about themselves.}

A plethora of HCI work has called for knowledge-based autonomy in one way or another.~\citet{DeVito2022-ee} poses folk theorization as a tool for autonomy in navigating hostile algorithmic systems. Regarding knowledge production,~\citet{Ajmani2024-ll} describe how the lack of trans autonomy over online content moderation contributes to epistemic injustice. In research,~\citet{C-Edenfield2021-by} speaks to leveraging the narrative expertise of trans research participants. We build on these calls and explicitly orient HCI research towards autonomy. Specifically, we focus on autonomy in knowledge settings because research is a form of knowledge production. Our call for epistemic autonomy is not exclusively transfeminist. \edit{For example, research that affords a community autonomy over their own beliefs is a key value in standpoint theory~\cite{Harding2001-sj}}. Still, given the stake transfems have in the concept of autonomy, we use transfeminine experiences and transfeminist thinking as our touchstone. However, we lay out principles that go far beyond the transfem-specific.

\subsection{Research Paradigms}
A research paradigm is a theoretical toolkit for articulating the beliefs that inform our actions. We distinguish the core beliefs of popular research paradigms. \citet{Guba1994-fj} define paradigms as ``\textit{basic belief systems based on ontological, epistemological, and methodological assumptions.}'' To that extent, when we take a paradigm-first approach to formulating and answering research questions, we can ensure that we address how our assumptions affect the research outcomes.


To illustrate the importance of having an explicit research paradigm,~\citet{Berryman2019-qb} presented The Cooking Metaphor. To paraphrase, imagine a world-class kitchen. This kitchen is where delicious foods are prepared by a chef who is passionate about their food and the people who will eat it. This chef carefully chooses recipes, ingredients, and cooking techniques to create good food. \textit{What's simmering on the stove?} The upshot is that despite our assurances that the chef is equipped, passionate, and well-intentioned, we do not know what they consider good food. For example, a vegetarian chef may have the same kitchen as an omnivore but use different ingredients.

Now, imagine someone asks you to ``do ethical research.'' Like a chef, you need to choose core beliefs: a set of recipes, ingredients, and cooking techniques that will likely create good food. When we start from an express research paradigm, we choose our core ontological, epistemological, and methodological beliefs.

\textbf{Ontology} is the study of reality and, therefore, the foundational layer of a research paradigm. Reality outlines what \textit{can} be known. In research, suppose we take a strictly \textit{positivist} ontological position: there is a single objective reality~\cite{Bridges2014-ep}. This would entail that there is a single set of truths that we uncover through scientific inquiry. Everything beyond this can be understood but not truly known.

\textbf{Epistemology} builds on ontology to study the ways we come to know things. Ontology is the nature of reality, and epistemology is the nature of knowledge formation. Epistemic questions are often relational between the researcher (i.e., knower) and information (i.e., would-be known). For example, feminist scholar Donna~\citet{Haraway1988-ch} proposed the epistemic theory of ``situated knowledge,'' it is impossible to separate knowledge from the context in which it was created, including who the knower is.

\textbf{Methodology} is the ideological umbrella that encompasses research design, methods, approaches, and techniques. For example, a positivist methodological approach could entail data collection methods, causal analyses, and internal validation techniques. 

HCI is a unique field in that multiple paradigms can co-exist, and individual paradigms can intersect.~\citet{Bodker2006-vy} describes the chronological shifts in HCI beliefs as it moved from an engineering discipline toward the social sciences.~\citet{Harrison2007-hr} outline three paradigms of HCI that all exist at the same time. Notably, HCI is filled with subfields that lie at paradigmatic intersections. For example, human-centered machine learning~\cite{Chancellor2023-gd} intersects the positivist views of developing valid measures of model accuracy, fairness, and bias with the critical view that model development is subject to socially constructed forces of power and identity. In this paper, we propose a research paradigm that practitioners can adopt to center participant knowledge. This paradigm may need to intersect with others to create useful outcomes. We argue that epistemic autonomy as a research paradigm encourages CHI's self-stated goals of research justice~\cite{Collective2021-cg, Chordia2024-ff}, participatory efforts~\cite{Longdon2024-kl, Pierre2021-zz, harrington2019deconstructing, Cooper2024-xg}, and fighting marginalization~\cite{Winschiers-Theophilus2013-yz, Bardzell2011-ln, Ogbonnaya-Ogburu2020-wt}. However, we do not suggest that our paradigm is ubiquitously better for all CHI research projects and, by proxy, HCI. In fact, the beauty of research paradigms is that they are tools researchers can have at their disposal and decide to use where appropriate.

\section{Establishing Epistemic Autonomy in Research}
In this work, we adopt autonomy from the transfeminist ideal of bodily autonomy---one's ability to govern the fate of one's own body and life without coercion or violence~\cite{Hines2019-fj, Koyama2020-zi}. In this section, we establish \textbf{epistemic autonomy}---one's ability to govern how they know themselves and how they are known to the world. In the same way that bodily autonomy is a fundamental right that is disproportionately taken away from transfems, so is epistemic autonomy. We argue that HCI research is a key site of violations of epistemic autonomy. At the same time, HCI research is uniquely poised to afford epistemic autonomy through a paradigmatic shift in our current methods.


In the following sections, we establish two forms of epistemic autonomy: (1) testimonial and (2) hermeneutical. Starting from the definitions of testimonial and hermeneutical injustice articulated by~\citet{Fricker2007-zh}, we argue that incorporating the transfeminist ideal of autonomy forces us to rethink the priors of current HCI research.

\subsection{Testimonial Autonomy}
Inspired by testimonial injustice, we define \textbf{testimonial autonomy} as one's ability to govern their credibility. Testimonial autonomy is the absence of prejudices that delegitimize one's authority, credentials, and expertise. For example, a common testimonial injustice is not believing transgender folks as credible authorities on their own gender identity. In an environment with testimonial autonomy, people are given the right to testify their own gender identity without question. Moreover, in the same way bodily autonomy demands freedom from coercion or violence, testimonial autonomy demands freedom from attack. In other words, testimonial autonomy is the intersection of testimonial justice and safety---the ability for one's testimony to be believed and not attacked. 

In an HCI context, we are explicitly concerned with the testimonial autonomy of research participants from marginalized communities. Specifically, how does our research violate testimonial autonomy? Considering this question requires us to reflect on our beliefs about participant testimony. In this paper, we highlight violations of transfeminine testimony. We present stories highlighting how statements of transfems are consistently discounted both in our sites of study and through our research practices (see Section~\ref{sec:violations}). We argue that these violations will continue without a new set of ontological, epistemological, and methodological commitments (that is, a new research paradigm) that centers epistemic autonomy.

\subsection{Hermeneutical Autonomy}
We define \textbf{hermeneutical autonomy} as an individual's ability to govern the sensemaking terms and concepts (hermeneutic resources) used by themselves and others to describe their experience. Recall that hermeneutical injustice is the systemic denial of sensemaking resources around one's experience~\cite{Fricker2007-zh}. As people move through the world, build community, and share experiences with others, they create hermeneutic resources.

Transfeminism argues that community-specific terms, such as gender identities, are powerful sensemaking resources that can combat the isolation and stigma transfems currently face~\cite{Koyama2020-zi}. Without concepts such as transgender, gender dysphoria, and transmisogyny, it is difficult for trans folks to find community with each other~\cite{Jackson2018-io}. This isolation leads to material harm for the trans community, as an inability to communicate one's experience and find support contributed to the disproportionate suicidality rate among trans folks~\cite{McNeil2017-by, Moody2015-ry}. This phenomenon of hermeneutical injustice contributing to systemic psychological and physical harm exists in numerous other marginalized communities, such as Black~\cite{Hull2017-fc}, Indigenous~\cite{Tsosie2017-bp}, and trauma survivor communities~\cite{Jackson2019-rp, Medina2012-dn}. 

In that way, everyone has a fundamental right to label their experiences as they see fit and not use labels that do not resonate with them. In HCI research, hermeneutical autonomy involves reflecting on the terms and concepts we use to describe the communities we study. HCI research is a translational effort---we translate human experiences of interacting with technology into takeaways, papers, and calls to action. However, poor translation from the vocabulary of our participants to the vocabulary of research risks devaluing the hermeneutical resources our participants use~\cite{Ajmani2024-ll}. In essence, this lack of hermeneutical autonomy signals to participants that they are worthy of being observed but not worthy of labeling their observations and experiences. Therefore, the starting place must be the vocabulary of the humans and communities we study. If we start by constructing knowledge using research terms, we disempower research participants. We implicitly tell them that the words and concepts they use to interpret their experiences are wrong. Hermeneutical autonomy demands that we start from a vocabulary that is \textit{of the community} rather than \textit{of the research field.}

\subsection{Related Efforts in HCI}
\edit{Our calls for epistemic autonomy build on an ongoing trend in HCI to participate with our design audience and empower those in the margins. For example, the Scandinavian tradition of design influenced by worker rights and democracy inspired early participatory design work in HCI~\cite{Muller1993-ku}. Since then, HCI has adopted different instantiations of participatory design that reimagine participation, such as co-design~\cite{Vines2013-gj} and value-sensitive design~\cite{Friedman1996-kx}. Most relevant to epistemic autonomy, participatory action research approaches rely heavily on community and participant voices. Participatory action research rests on the epistemology of situated knowledge~\cite{Haraway1988-ch}; knowledge is contextual based on lived experiences. Especially within these participatory paradigms, HCI methods often center on giving voice to participants.}

\edit{We propose epistemic autonomy as the next step for empowering participants. Though successful, participatory methods are epistemically nuanced and, at worst, can further marginalize. For example,~\citet{Pierre2021-zz} describe participatory design as epistemically burdensome for minority groups. In addition to research participation, these communities are tasked with educating privileged researchers on their experiences of marginalization. Overall, participatory methods in HCI have long grappled with the role of the designer versus participant~\cite{Kensing1998-gk, Vines2013-gj}. To address these concerns, we contribute epistemic autonomy as a mechanism for empowering research participants as \textit{knowers.} Specifically, epistemic autonomy asks: can we---as researchers---respect the knowledge created by our participants and sites of study?}


\section{Researcher Positionality}
The first author of this paper is a cisgender woman with South Asian heritage. The second and third authors are transgender women and elaborate on their positionalities in the context of their personal stories presented in Section~\ref{sec:violations}. As a team, we are deeply versed in the literature of epistemic injustice with the first author leading significant previous work on the topic. The first author recognizes that, as a cisgender woman, she can only understand the epistemic injustice of trans women secondhand. Therefore, this work is intended to bring the perspectives, positionalities, and experiences of the second and third authors to bear on HCI research. 

\section{Stories and Violations of Epistemic Autonomy}\label{sec:violations}
In this section, we use storytelling methods~\cite{Ogbonnaya-Ogburu2020-wt} to articulate the harms of violating one's epistemic autonomy. Above, we articulate epistemic injustice and autonomy as theories, but the damage caused by ignoring these theories is real. We present stories from the two transfeminine authors of this paper to highlight two different contexts. First, an author who has moderated and maintained multiple transfem online communities speaks to the epistemic injustice she experienced over nearly a decade of community work, all pointing to HCI-relevant problems with unexpected dynamics (Section~\ref{sec:stories1}). We highlight these stories because, without understanding these experiences of epistemic lack of autonomy, HCI risks propagating the epistemic injustice happening in our sites of study. Second, a professor who researches the trans experience in digital settings speaks to her experiences of trying to fight epistemic injustice within HCI (Section~\ref{sec:stories2}). We highlight these stories because research can be a site of epistemic injustice. Our research practices are primed to violate the epistemic autonomy of our research participants. Therefore, it is crucial we interrogate our current research beliefs and progress them to empower those in the margins.

\subsection{Epistemic Autonomy in Online Communities: Intersecting Gender and Race}\label{sec:stories1}

In the following stories, I will draw mainly from my experiences online as an Indian trans woman in three different asynchronous communities:
\begin{enumerate}
    \item A private Facebook community in the late 2010s, run by and for queer Asians. I was appointed admin a short while after the community was formed.
    \item A private Slack community in the mid-2010s, run by former fans of the online feminist magazine The Toast. I was not an official community leader, but I was one of the very few trans women present and the only trans woman of color from the Third World. I was granted a degree of “authority” over various issues where different Western and white people found their knowledge lacking.
    \item A private transfeminist Discord server that I began, which I eventually had to delete and leave due to excessive demands placed upon me coupled with unfounded accusations.
\end{enumerate}
In each case, it bears mentioning that, in addition to gender, race played a role. As intersectional feminism articulates~\cite{Collins2020-bx}, my marginalized identities, such as my identity as a trans woman, compounded to create an easy target for others---erasing my vulnerability and positionality in the process.

\subsubsection{Story 1. Intracommunity Epistemic Injustice on Facebook}
For a brief period in the late 2010s, the Facebook group ``Subtle Asian Things'' kicked off a wave of similarly-themed private groups centering around Asian identity. I was part of a spin-off group called ``Queer Asian Intersections.'' I was not a member of the initial moderator team, but I quickly became an admin due to my activity within the group.

The moderation team was fairly tight-knit. Mods came from various backgrounds, encompassing South, East, and South-East Asians. There were a good mix of bi, lesbian, gay, transmasc, and transfem moderators as well, so QuAInt was unique in not making me feel like I was the only transfem voice in the room.

The biggest issue with the community was the classic fault line in Western, Asian diasporic politics: the question of assimilation, class, and national identity contrasted against origin and heritage. Despite being racially marginalized, many of the group’s members were from the middle classes, and were prone to reproducing hegemonic racial discourses about other racially marginalized demographics. There was a degree of internal tension as well since colorism\edit{\footnote{\edit{Colorism is a form of discrimination based solely on skin tone rather than race. Colorsim is particularly common in South Asian cultures where lighter skin is often seen as ``better than'' dark skin~\cite{Banks2015-lm}.}}} and racism against other Asian groups were rampant. These racial power structures intersected with my gender as a trans woman and pushed me into the margins, despite my authority as a moderator.

\edit{I was only given authority to moderate members who were also gender minorities. Otherwise, I received pushback from the cisgender moderators on what was considered transmisogynistic. I recall a particularly odd discourse where a large contingent of queer members ruled out dating trans partners due to our inability to reproduce. In addition to the explicitly transphobic arguments members made, these statements were a contradiction with their own queer identities. These contradictions clearly did not register. I raised the transphobic roots of this conversation to other moderators, but I was brushed off. A singular Black-Asian moderator was contorted into a position similar to mine. She was also viewed as the sole representative and authority on various issues, but had no authority over the racially charged and colorist comments other members made.} This was a bizarre form of testimonial injustice, where we were only given credibility to speak on behalf of our communities as a monolith rather than on behalf of our personal experiences as queer women of color.

This was a period of my life where I would not consider myself particularly well-read on transfeminism or transmisogyny. In hindsight, I can identify my core issues with the group as a result of the transmisogynistic and colorist culture. The queer community has long dealt with internal transmisogynistic and transemasculative discourses~\cite{Weiss2012-va, Scheuerman2018-uo} \edit{and this group was no exception.}  Eventually, I turned in my resignation and left the group.

I find this a notable case of injustice because even in the utter absence of white interlocutors, racial discourses converged around and reproduced white hegemony. Solidarity only extended so far and did not spare me or others from intracommunity transmisogyny despite my position of ``authority'' as a moderator.

\subsubsection{Story 2. Weaponized Credibility in a Slack Community}
While I did not have official mod power in this Slack space, I was looked to as an authority on several matters in a white-majority space. I was one of the only trans women present, one of the very few lesbians, and certainly the only trans woman of color from the Third World. I was frequently instrumentalized in arguments and subsequently subject to a high degree of moderator action. A particularly galling incident involved a debate around Sarah Schulman's \textit{Conflict is Not Abuse}. I expressed discomfort with a majority-white space arguing for the denial of epistemic authority to marginalized members; much of the book's rhetoric seemed to advocate for not taking accusations of harm seriously. I was, ironically, disciplined and subjected to moderator action for being ``too heated.'' The irony was lost on everyone but me and other members of color. 

There was a policy of ``POC moderating POC''---an edict that ultimately did little to ameliorate white grievances. Subsequently, there was a pattern of appointing cisgender Desi moderators as a neutral epistemic authority over my experiences. However, they were not neutral as their moderation decisions were only respected if it was in agreement with the dominant Western perspectives. In this situation, none of us were given due credibility as Desi women, which compounded for me as the sole trans woman of color. Our credibility and authority was not self-governed, it was afforded and taken away by the white members of the community so that people of color were acting against each other.

\subsubsection{Story 3. Attack on a Self-Owned Discord Server}
I started a server on Discord that was initially meant to be a space for fans of my writing. Over time, the space developed, grew, and became a more transfeminism-focused space oriented around radical feminist theory. My responsibilities and duties compounded as the sole admin, and I was expected to treat a personal space as a movement space without much support from other community members.

The eventual implosion manifested as accusations of transmisogynoir---the specific oppression of Black trans women~\cite{Krell2017-iv}. Notably, these accusations did not come from any of the Black trans women on the Discord. I was made to answer for shortcomings of the space, both real and imagined. The accusations towards me spiraled into outright falsehoods quickly. Allegations of ``harboring ex-TERFs\edit{\footnote{\edit{TERF stands for a trans-exclusionary radical feminist. TERFs are a group of women who actively reject the inclusion of transfems into women's spaces~\cite{Pearce2020-qr}.}}},'' ``anti-transmasculinity,'' and ``Nazi sympathizing'' were supported by decontextualized screenshots. 

Through it all, I was denied the right to defend myself because, as the space’s sole admin, I was somehow responsible for the actions of its white members. I was also treated as though my own marginalizations and positionalities as a trans lesbian of color from the Third World ceased to be important. \edit{For example, the attacking members used my transness to discount my ability to properly administrate my own Discord server. My own struggles with the group’s white members were erased in order to position me as an oppressor who must be accountable for everyone's transgressions rather than just my own.} I was effectively deracialized, degendered, and disallowed the particularities of my personhood, all to enable the abuse of me, the burning of me in effigy.

\subsection{Epistemic Autonomy in HCI Research: The Trans Woman In The Room}\label{sec:stories2}



I am a transgender woman who has been publishing in HCI and social computing for slightly under a decade, and as I have moved from brand-new PhD student to Assistant Professor and PI of my own lab, I have seen a number of cases where centering epistemic autonomy as a research value would have made a substantive difference. Here, I will recount three key examples: one where I should have done more as the trans woman in the room; one where I had to step in as a more senior transfeminine voice; and one where I (hope) I did right by my own epistemic authority, and the epistemic authority of the trans women my work was for and about.

\subsubsection{Story 4. Propagating Cultural Erasure Through Research} When I was in grad school and very early in my own transition, I was called upon by colleagues to join a project on how Hijra\footnote{\url{https://taliabhattwrites.substack.com/p/the-third-sex}} in Bangladesh use technology. The project~\cite{Authors2000-ev} was extending one of my frameworks outside of a Western context for the first time. I was fascinated by both the theoretical progress we would make and by the notion of working on research about Hijra, who have classically been thought of as a group that has parallels with, but is quite distinct from, transgender women. Therein lies the problem, though: I am not a Hijra, and I was not on the ground in Dhaka, so my concept of Hijra had to flow from past work. If an academic asks to know more about Hijra, they will inevitably be directed to Serena Nanda's ``Neither Man nor Woman: The Hijras of India,'' which has been regarded as the canonical text on the subject for the last 24 years. I wanted context, so I went to the canon---but here, the canon is written by a cisgender woman who was born in and largely lived her life within New York City. Nanda is not and has never been a Hijra; she is not even transgender. Yet, she is the authority, so she informed our work. 

I now see this as a failure on my part, because in the years since, I have had the pleasure and the privilege of interactions and friendships with Desi trans women, who have educated me to the reality: many people labeled Hijra \textit{do} see themselves as transgender women. Hijra people advocate for themselves on that basis, and do not wish to be forcefully third-sexed as Nanda (and the oppressive local legislation based on that model of third sexing) insist is necessary. Reading Nanda's text now, I must reckon with a research decision I deeply regret. I regret citing it, as it goes out of its way to ignore the testimony of Hijra themselves in favor of rank anthropological Orientalism. I thought I was doing the right thing, but I let my status as a trans woman and the rest of the team's status as Desi people convince me that we understood the full cultural dynamics. But we did not respect the epistemic authority of the Hijra, and as a result, we did not even get the chance to consider the nuance that their own diverse understandings of themselves and their needs might have revealed. 

It is not sufficient to have a trans person or a person from the same culture substitute in as the epistemic authority. It needed to be a Desi trans woman making that call, preferably a Hijra herself, to do right by the participants. We can not simply cite canonical texts when that text explicitly robs the subjects of their own epistemic authority. We need to go to the people who \textit{rightly} hold that authority themselves if we really wish to understand the nuances of how people use technology in different cultures. 

\subsubsection{Story 5. Transfeminine Erasure within Trans Research} When I was a postdoc and much more settled in my identity as a transgender woman, I helped design a pair of projects with a cisgender colleague. Both projects included a heavy focus on transgender topics. My colleague ran the project on how transgender people experience and navigate eating disorder-related spaces online~\cite{Authors2000-eh}. Though she is not transgender herself, she is by far one of our best researchers on the topic of ED-related technology. At one point during the writing and editing phase of the project, when I had shifted much of my focus to running the other project, she noticed something: the results about trans women, specifically, kept getting left out of the evolving narrative. The other authors argued that there just was not a clear narrative around the transfeminine results, and the results for transmasculine participants were more interesting. Despite the fact that there were other trans people on the team, the transfeminine results simply were not meshing like the more transmasculine-focused components were. 

Over the course of an afternoon, I---with the positionality of a trans woman dealing with my own relationship to cis feminine beauty and body norms---was able to work with the first author and completely re-cast the transfeminine findings as a key point of contrast with the experiences of the transmasculine people the narrative already focused on. Instead of essentially abandoning half the trans participants because we could not find an obvious transfeminine narrative, which would have simply resulted in more testimonial injustice for the trans women we interviewed, I was able to use my own hermeneutic resources to help draw out what became the focus of the paper. Ultimately, the paper focused on the differential distances transfeminine and transmasculine people hold from the cis-feminine ideal and the need to build solutions and structures that account for, instead of collapse or ignore these key differences. Simply put, sometimes you need a trans woman with the appropriate epistemic authority (and the in-group seniority to back it up) to make sure work that spans multiple trans groups maintains epistemic justice for them all.

\subsubsection{Story 6. Affording Epistemic Autonomy From the Margins First}
The other project of those two I co-designed aimed squarely at one of my own problems: dating applications are terrible for lesbians, especially lesbians who also happen to be trans women, such as myself ~\cite{Authors2000-bp}. Here, I am willing to cautiously claim that I did things right, because this was the first project I ever designed with proper epistemic autonomy in mind. This could not be a project that only involved trans women because that is not how lesbian dating works. Lesbian trans women are part of the general pool of lesbians. To cut us off from the cis lesbian population would be an act of discriminatory violence and hermeneutical injustice in itself. Lesbian culture and lesbian traditions belong to all lesbians, not just cis lesbians. However, it is transgender lesbians, especially Black trans women, who are most harmed and excluded by the current apps. I was determined to focus squarely on those issues. As such, the design explicitly asserted and codified the lines of epistemic authority via rules: (1) discussion would prioritize the needs of trans women generally and Black trans women specifically; (2) when making design and policy decisions, the needs of trans women and especially Black trans women would come first. I was concerned that the cis women in the study would object, but to my pleasant surprise, not one of them did. Because the epistemic authority of trans women in the space was not only a high-level goal but an operational, enforceable principle of the research design itself, all participants respected and embraced that authority. These values carried over into analysis and decision-making during writing. We held to our rules, even when we might have personally made other choices, and maintained the epistemic authority of transfeminine participants throughout, backed up and contextualized by a largely-transfeminine research team. The results are a paper I am extremely proud of, full of design directions that I am confident will actually keep trans women safer as trans women do one of the most basic positive things most humans do: seek loving companionship. It was the first thing I ever wrote that I was confident in calling a ``transfeminist'' work - another explicit signal of how seriously we took the epistemic authority of trans women. 

We need to take the epistemic authority of the most marginalized people we design for and work with seriously in research, or else we miss nuance and erase whole narratives. This does not happen through any kind of malice. We (researchers) are not scheming villains, but are busy people who simply \textit{do not have} the relevant experience or context to realize they've committed or furthered epistemic injustice. 



\subsection{Themes of Epistemic Injustice and Autonomy}
The above stories describe numerous instances of epistemic injustice and, thereby, lack of autonomy. Below, we describe the central themes from these six stories and how they evidence a path toward epistemic autonomy as an antidote to injustice. Story identifiers are abbreviated as S\#.
\subsubsection{Testimonial Injustice: Credibility Governed by Someone Else}
Recall that testimonial injustice occurs when an individual is not given due credibility because of prejudices against them. A straightforward example of testimonial injustice is when transgender folks are not believed to be credible testifiers to their own gender identity. The stories above highlight a more nuanced form of testimonial injustice where your credibility is governed by someone else, particularly someone in a dominant position. For example, S2 describes how people of color in an online community were mainly given moderator authority to moderate other people of color. Moreover, moderator decisions were not given due consideration if they did not affirm the dominant position. S5 describes a similar scenario where transfem representation within a research paper was governed by the projects researchers and, because of this, were nearly erased entirely. Fortunately, an transfem author was brought on the project and prevented this erasure. Researcher interpretation is a crucial component of HCI methods. However, it is a ripe venue for perpetrating epistemic injustice by systemically erasing participant voices within an identity group. The solution here is to commit to testimonial autonomy---the self-governance of credibility and authority of our research participants. 

\subsubsection{Hermeneutical Injustice: Queerness \& Transness without Transfems}
Hermeneutical injustice occurs when one is systemically stripped of the necessary resources to make sense of their experience. Identity labels such as ``queer'' and ``trans'' are powerful hermeneutical resources. Queer spaces where people share their experiences and build community are empowering, especially given the history of queer folks being forced into invisibility (e.g., Don't Ask, Don't Tell~\cite{Human-Rights-Campaign2010-ht}). Yet, the above stories highlight how queerness and transness are currently being weaponized and misappropriated. For example, in S1 the author describes how her own queer and Asian community was othering trans folks due to their inability to reproduce, despite many queer cisgender couples also not being able to reproduce. In this case, the hermeneutical resource of ``trans'' was wielded as a weapon against people within the same online space. On the other hand, S4 highlights how transness was not justly afforded to the Hijra community. In this case, the trans label was taken away from the Hijra community by a researcher and writer who was not Hijra herself. The Hijra identity and their transness were defined by others.~\citet{Falbo2022-jy} describes these forms of hermeneutical injustice as conceptual distortion. Rather than there being an absence of hermeneutical resources, the resources are owned by others and, therefore, used to disempower marginalized communities. We argue that the antidote here is hermeneutical autonomy---ensuring the terms we use to describe communities in our research are of the community.

\subsubsection{Affording Epistemic Autonomy}
Our stories demonstrate how centering epistemic autonomy is a useful mechanism for fighting epistemic injustice. S5 and S6 articulate how researchers can center the epistemic autonomy of their participants. Specifically, S6 articulates the explicit research decisions needed to first afford epistemic autonomy and authority to those in the margins. To respect intersecting identities, the researcher included both cis and trans lesbians in the study. To maintain testimonial autonomy, the researchers set aside their own views for the sake of the participants. As articulated in S1-3, our sites of study often have their dynamics of gender and race that compound to epistemic injustice. Even when she started her own community, the author was subject to systemic silencing of her experiences as a trans woman of color (S3). However, S6 shows that proactive research practices have the power to stop and potentially remediate the epistemic injustice seen in our sites of study. By centering epistemic autonomy, particularly of those who are often victims of epistemic injustice, researchers can start to create more just online environments, discourses, and research outcomes.


\section{Epistemic Autonomy as a Research Paradigm}
\begin{figure}
    \centering
    \includegraphics[scale=0.3]{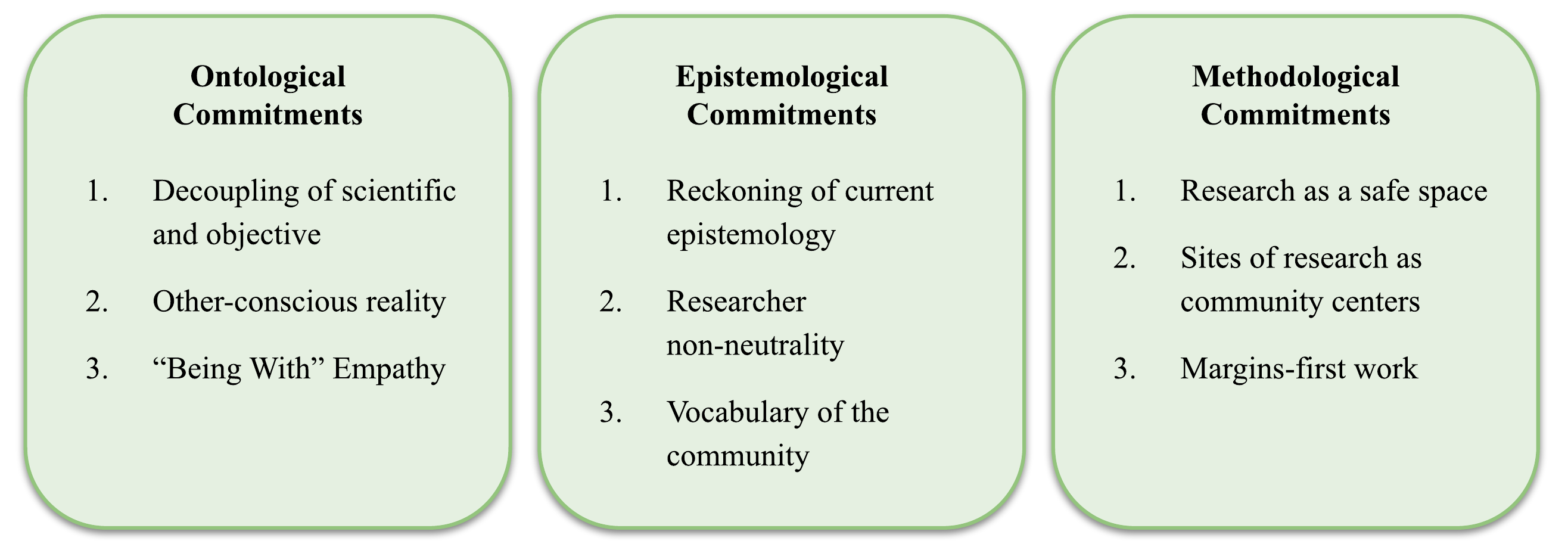}
    \Description{Three-stage diagram of the ontological, epistemological, and methodological commitments of epistemic autonomy as a research paradigm}
    \caption{Summary of commitments for affording epistemic autonomy to participants in HCI research.}
    \label{fig:paradigm}
\end{figure}

Research paradigms are the set of core beliefs that guide research processes. For example, constructivist and interpretivist paradigms hold that reality is created by individuals. Therefore, the process of finding \textit{``true knowledge''} (i.e., doing research) is the process of interpreting the individual experience and discovering underlying meaning. Having explicit research paradigms is crucial to understanding the ethical implications of our methods. For example,~\citet{Bardzell2011-ln} articulate how---without an explicitly feminist paradigm---HCI risks propagating social and structural inequities. In this section, we outline the paradigm shift that needs to happen in HCI to make room for epistemic autonomy (summarized in Figure~\ref{fig:paradigm}). While we name an epistemic problem here (i.e., epistemic injustice), the solution requires a fundamental change in our ontological and methodological beliefs. To that end, we highlight nine commitments that HCI as a field must make to stop the propagation and further creation of epistemic injustice. Without this paradigm shift, HCI risks contributing to the systemic violence that marginalized communities, such as transfems, face. With this shift, we can reveal the power of HCI research as an antidote to epistemic injustice.



\subsection{Ontology. One's Experience is their Reality}
At the heart of epistemic injustice is disbelief---disbelief that someone with identity X can know Y. Disbelief about someone's lived experience presumes that experience and reality are two distinct entities. Therefore, someone's experience can be devoid of reality and not merit respect. Ontologically, we must stop attempting to disentangle experience from reality. Here, we outline the ontological commitments (i.e., presumptions about reality) that HCI researchers must make. Without these priors, there is little space for marginalized folks, such as trans women, to gain epistemic autonomy. 

\subsubsection{Decoupling of Scientific and Objective.} Formally, objectivity is the belief that an object can exist independent of one's perception~\cite{Daston2021-im, harding1995strong}. In effect, there is a reality of objects that exists regardless of human perception, experience, and social behavior. Objective claims are viewed as free from human biases and, therefore, factual. HCI is a movement away from objective agendas in computational sciences and towards human-centered ones~\cite{Myers1998-qk, Ackerman2000-ik}. This move relies on the fundamental premise that technology has implications beyond the objective. Technology can be political~\cite{Ghoshal2020-xz, Kasirzadeh2022-vr}, personal~\cite{DeVito2018-gr}, and social~\cite{Shelby2023-ig}. However, technology is just our object of scientific study. Solely critiquing technology still prioritizes monolithic takeaways over individual testimony~\cite{Ajmani2024-qk}. 

An explicitly autonomous approach builds on the movement to critique technology and towards critiquing scientific study itself. If the mission of HCI is to center the multitudes of human experience, why are we operating from a presumption of a single scientific way? Studying how people use technology demands rejecting objectivity. Unlike natural sciences (e.g., physics, chemistry), our entire field would not exist independent of the human experience. Therefore, it is inappropriate for us to center those presumptions of scientific reality.

\subsubsection{Other-Conscious Reality. }Constructivist and critical ontologies imply that there are multiple, ever-evolving realities that are constantly under social influence. The flip side of this belief is that research, which seeks to uncover reality, has the power to influence.~\citet{Ogbonnaya-Ogburu2020-wt} describe the act of considering this influence as being ``other-conscious.'' Being other-conscious involves considering experiences beyond one's own. We frame this as an ontological belief because it concerns the realities that researchers are prone to accept versus the ones we choose to disregard. For example, we often consider the effect our research will have on senior researchers who may review our papers but fail to consider how racial minorities may be affected by our research~\cite{Ogbonnaya-Ogburu2020-wt}. This idea of other-consciousness is echoed in transfeminine literature~\cite{Koyama2020-zi}, where there is an explicit call for transfeminism to consider other philosophies with intersecting axes of marginalization, such as Black~\cite{Taylor2017-vm} and Indigenous feminism~\cite{Green2020-yz}. Being other-conscious is the foundation of affording epistemic autonomy. How can we give people space to govern knowledge without also considering how our research affects them?

\subsubsection{``Being With'' Empathy. }When we accept experience and reality as one and the same, we remove researchers from the role of ``reality discoverer'' and into the role of ``experience hearer.'' In other words, researchers are tasked with empathy towards participants. \citet{Bardzell2011-ln} advocate for an increased focus on having an \textit{``empathetic relationship with research participants.''} In practice, empathy as a research approach has been a mixed bag, as it has sometimes been used as a cover for approaches where we practice ``being like''---a focused form of empathy where we attempt to understand other people through simulation and roleplay instead of direct engagement. True empathy, as~\citet{Bennett2019-gf} have described in their work from the accessibility subfield, requires a ``being with'' approach, where researchers attempt to attune their understanding with participants on a regular basis instead of engaging once and trying to simulate from there on out. \edit{\citet{Spiel2024-pu} describes this as practicing design humility, which involves expanding a designer's lens instead of overriding other's experiences to fit through this lens.} In other words, a participant's experience is their reality.

To pretend that we can judge the ``realness'' of an experience that we ourselves have not had is to engage in testimonial injustice. We should instead be focused on the testimony of those with real-life experience, holding ourselves to their experience instead of our internal guesswork. 
\subsection{Epistemology. Epistemic Autonomy}
Epistemic autonomy is the capacity for one to self-govern their understanding of their own experience. In keeping with epistemic injustice theory~\cite{Fricker2007-zh}, we argue that being able to know your experience is a fundamental right. In this section, we outline three epistemic commitments HCI researchers must make to give autonomy to research participants.

\subsubsection{Reckoning of Current Epistemology. }
Epistemic autonomy begins with a reckoning of our current epistemology. Epistemologies themselves are often derived from and function as tools of patriarchy~\cite{Fricker2007-zh, Haraway1988-ch, Menking2021-ic}. We must ramp up our interrogation of positivist thinking and push for the widespread acceptance of epistemologies that directly reflect the fact that all knowledge is socially constructed and situationally contextualized~\cite{Haraway1988-ch}. We must actively combat the privileging of quantitative measurement over qualitative inquiry and recognize that the overwhelming drive to quantify is, in fact, an act of marginalization. It does not matter if some experiences only represent a small fraction of the entire world’s experiences---they still exist. To ignore or devalue these experiences simply because they do not scale or validly measure is an act of epistemic injustice. Time and time again, the minority status of transfems and trans people overall has been used as a bludgeon: an excuse to exclude an entire community from studies, an excuse to collapse all trans experience down into one category, an excuse to not fund and not prioritize the study of trans people. The drive to quantify and the related drive to force all knowledge through the lens of positivism are both engines of the very epistemic injustice we must fight. 

\subsubsection{Researcher Non-Neutrality.} Research neutrality has been a longstanding controversy in the social sciences~\cite{Harding1992-xw}. Durkheim championed positivist and objective social sciences, arguing that a sociologist can be neutral~\cite{Durkheim2023-jk}. However, his critics---such as Marx---argue that Durkheim’s call for objectivity hides an ideological preference for the existing social and political order~\cite{Bottomore1981-th}. For example, the moral imperative to improve the world through research is incompatible with a neutral researcher. A goal to improve the world is morally non-neutral. In other words, to be objective is to be complicit in the status quo.

Similarly, HCI has embraced the idea of researcher positionality~\cite{Liang2021-qb}. The epistemic belief here is that a researcher has identities, relationships, and positions that influence the knowledge they are claiming to create. Therefore, HCI researchers are encouraged to self-disclose their stances that could have affected the work. However, encouraging self-disclosure still presumes that neutral research is ideal, and positionality disclosures help us move closer to this ideal by accounting for non-neutral stances. Our approach calls for a complete dissolution of ``researcher neutrality'' as a concept. Inspired by Gouldner's response to Durkheim and Marx, we argue that a researcher, as a human being, undergoes change through the research process~\cite{Gouldner1974-gx}. Therefore, researchers are participants of their own research. Similar to how we treat the experiences of research participants as non-neutral, we should do the same to our own. 

\subsubsection{Vobabulary of the Community.} HCI has a rich history of bringing communities into our generative activities. Most notably, participatory design is a longstanding and well-respected method in our field~\cite{Sanders2002-tn}. However, whose lenses are these communities participating through?~\citet{Fricker2007-zh} describes hermeneutical injustice as a system of oppression where communities are not allowed to create or adapt the vocabulary used to describe their experiences. Research, particularly in HCI, demands having a trained middle-man---the researcher---to interpret and communicate participant experiences. Therefore, research can potentially introduce terms that misappropriate our communities of study, as demonstrated with the Hijra community in S4. In keeping with our call for hermeneutical autonomy, failed interpretations of a community's vocabulary are epistemically invalid. Knowledge that uses flawed terminology is not true knowledge and, therefore, is not valid research.

\subsection{Methodology. Centering Participant Knowledge Free From Attack}
Methodological commitments are explicit duties researchers have during their studies. These commitments then inform the methods, analyses, and tools that researchers use to generate results. Affording epistemic autonomy involves centering participant knowledge and creating a space free from attack. Stemming from the transfeminine articulation of bodily autonomy, epistemic autonomy is free of coercive or threatening forces. In many venues used in HCI research (online communities, research workshops, etc.), harassment of marginalized voices is rampant. Epistemic autonomy holds that it is the researcher's duty to create safe spaces, respect the community, and work from the margins first. These commitments inform two methods articulated in Section~\ref{sec:application}.

\subsubsection{Research as a Safe Space}
While spotlighting marginalized communities through research is epistemic justice, it carries a large amount of researcher responsibility.~\citet{Ajmani2024-ll} specifically call this out as a tension of visibility in the fight for epistemic justice. As individuals and communities become more visible, they become larger targets for harassment and hate speech~\cite{Scheuerman2021-nj}. Given researchers are using these communities as sites of study, we are responsible for keeping them safe.

\subsubsection{Sites of Research as Centers of the Community.} 
This methodology demands considering communities as complex sociological structures rather than monoliths. This is especially important in spaces involving intersectional identities, such as transness and femininity. Here, we try to take epistemic authority away from folks partly by denying individual facets of their identity experience in order to devalue the rest. For example, Story 3 (see Section~\ref{sec:violations}) describes how the author's role as an administrator took away her right to call out transmisogynistic activity in an online community she started. When someone's experience looks different, it breaks down our narrow-minded definition of community. If community necessitates similarity, then there is no space for differences within a community. Often, these differences fall along identity lines, adding layers of marginalization to an already fraught space.

\subsubsection{Work from the Margins First. }Finally, we call for HCI to work from the margins first. At face value, this methodological commitment conflicts with designing usable, general, and popular systems. However, designing for the center inherently propagates marginalization. Recall how objectivity carries an implicit okayness with the status quo. Designing for general populations rather than specific communities carries a similar okayness with marginalization. Those who are in the margins will continuously be pushed into them. Our research is a tool that affords power. More often than not, we use it to empower those who already dominate.



\section{Applying Epistemic Autonomy to HCI Methods}\label{sec:application}
In this section, we apply our paradigm to two common methods in HCI: autoethnography and asynchronous remote communities (ARCs). We describe how applying epistemic autonomy gives us new flavors of these methods that actively promote the testimony and hermeneutic resources of those in intersecting margins.
\subsection{Scaffolded Autoethnography}
As its name suggests, autoethnography is a self-study research method where one reflects on, analyzes, and synthesizes their own experiences. Concisely, autoethnography is a self-reflective form of cultural analysis~\cite{Chang2016-bb}. In HCI, autoethnography has been used by researchers to highlight the unique tensions that arise in being a technology expert and a frustrated user~\cite{Lucero2018-gl, Fassl2023-ur}.~\citet{Erete2021-ul} leveraged this intersectional nature of autoethnography to describe the painful experiences of being continuously marginalized as Black researchers. Most relevant to this work,~\citet{Ajmani2024-ll} used a collaborative variant of autoethnography to raise personal testimonies of experienced and observed epistemic injustice in online communities.

\begin{figure}
    \centering
    \includegraphics[width=0.9\linewidth]{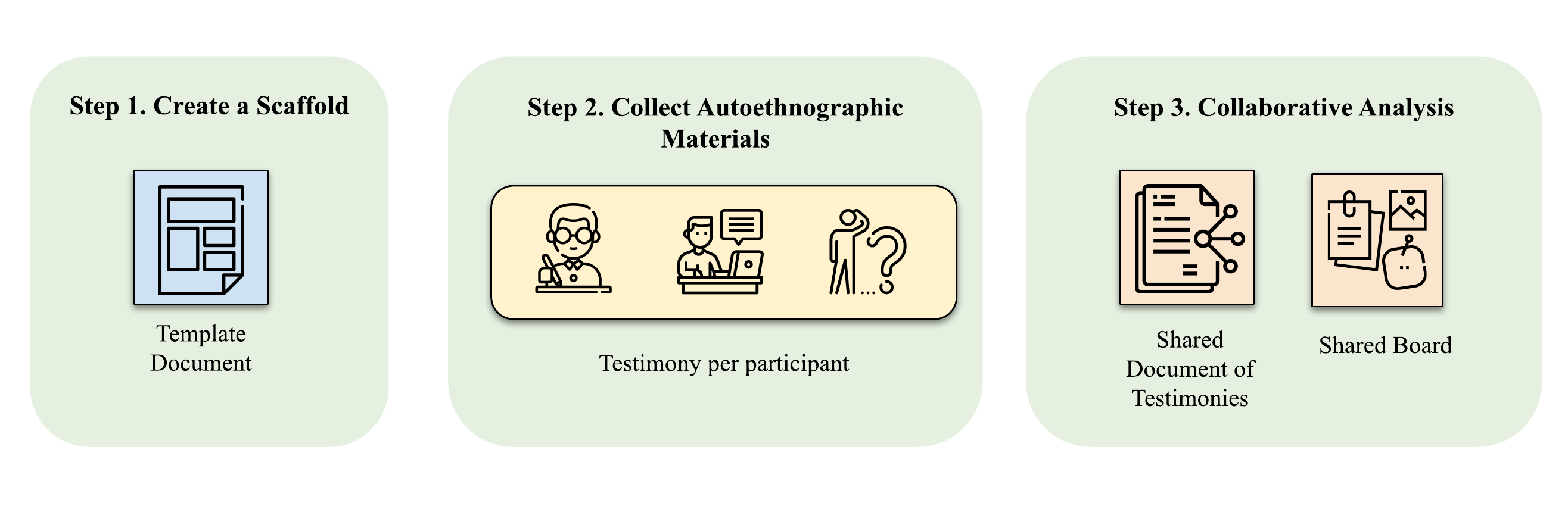}
    \Description{Three-stage diagram of the steps to implement a scaffolded autoethnography}
    \caption{Steps and associated documents for scaffolded autoethnography (SAE). Inspired by collaborative autoethnography, SAE involves facilitating participants to be autoethnographers by scaffolding their materials and facilitating collaborative analysis.}
    \label{fig:scaffolded-auto}
\end{figure}

While autoethnography has become a powerful method within HCI, it is still limited to self-study of trained researchers. We propose a method called \textbf{scaffolded autoethnography (SAE)} where research participants are trained to contribute autoethnographic materials, analyses, and research findings. \edit{Scaffolded autoethnography furthers epistemic autonomy by relying on participant testimony as data and involving participants in the analysis of said data.} Testimonially, SAE gives participants their due credibility as crucial contributors to our research findings. Hermeneutically, SAE encourages a vocabulary that is of the community and creates the opportunity for participant sensemaking. Below, we translate~\citet{Chang2016-bb}'s phases of collaborative autoethnography into a method of scaffolded autoethnography (see Figure~\ref{fig:scaffolded-auto}). Using the principle of collaborative autoethnography, SAE can be scaled to include multiple autoethnographers.

\subsubsection[]{Step 1. Creating the Scaffold}
In traditional autoethnographic methods, the researcher determines the topic and scope for themselves. To that end, they may create structures to organize the project topic and scope, such as reflection templates~\cite{Lucero2018-gl}, protocols~\cite{OKane2014-fa}, or an implicit set of rules~\cite{Fassl2023-ur}. SAE starts with creating a similar structure for participants. Originating from education settings, scaffolding here is the process of creating structures where the primary objective is for an expert to teach a novice something~\cite{Gordon-Wells1999-xn, Weissberg_Hyland_Hyland_2006}. \edit{This idea of a scaffold helps achieve epistemic autonomy of participants by bridging skills gaps between the researcher and the participant.} To achieve these objectives, we suggest researchers create documents that (1) outline the scope and topic of the research project and (2) teach participants autoethnography to center them as autoethnographers. We specifically use the term ``scaffold'' to highlight that these documents are intended to lightly scope autoethnographers rather than constrain. 

Specifically, a good scaffold (1) establishes a common goal for the research project, (2) provides scope to participants, and (3) gives participants leeway to contribute what they feel is relevant on their own terms. First, as~\citet{Chang2016-bb} articulate, setting a goal for autoethnography involves outlining the researcher's anticipated contribution and defining any fundamental concepts (e.g., epistemic autonomy). Second, scoping participants' autoethnographic materials requires a document with guides for participants. For example,~\cite{Ajmani2024-ll} provided autoethnographers with a mixture of headings and questions to appropriately scope their reflections. Third, participants are given instructions on how to engage with the scaffolding materials. A scaffold, such as a template, is distinct from providing a list of questions to be answered. Rather, the autoethnographers are encouraged to use or ignore the prompts as they see fit. This template is then circulated to potential autoethnographers to collect autoethnographic materials.

\subsubsection[]{Step 2. Collecting Autoethnographic Materials}
The data collection phase is the stage of collecting raw autoethnographic materials. In previous work, this is typically a combination of field notes, personal data, and retrosepctives~\cite{Lucero2018-gl}. To make autoethnography as accessible and unrestrained as possible, scaffolded autoethnography mainly focuses on testimony as data. Testimony is an individual recounting of a personal experience. \edit{Following the ideals of epistemic autonomy, participant testimony is itself a form of knowledge. These testimonies are spaces for participants to communicate their knowledge, experiences, and understandings.} To that end, Step 2 involves collecting individual testimony from every participant (i.e., autoethnographer). In Step 3, these testimonies get coalesced into a shared document with all autoethnographers.

\subsubsection[]{Step 3. Collaborative Analysis}
In its original form, collaborative autoethnography relies on collaborative analysis. Here, everyone who contributed an autoethnography is brought together to share, analyze, and interpret~\cite{Bala2023-mf}. In scaffolded autoethnography, analysis is this process of coming together for collective interpretation of testimonies from Step 2. Here, the researchers are facilitators and coaches rather than contributors. 

To start analysis, autoethnographers and researchers come together in a shared space (either virtual or in-person). With proper permission, all members are given access to (1) a shared document of testimonies and (2) a shared post-it note board. This method of having shared, visible resources that all participants can manipulate is inspired by social translucence~\cite{Erickson2000-au} and has been used in previous HCI collaborative autoethnographies~\cite{Ajmani2024-ll, Bala2023-mf}. For each testimony, autoethnographers take time to read the testimony in full and then collectively discuss important features of the individual case. During this discussion, the researcher serves as a scribe and documents the features identified as post-it notes on the shared and visible board. It is important to note here that ``features'' are not necessarily themes or commonalities across cases. In fact, a distinct use of collaborative autoethnography is to surface paradox and difference~\cite{Bala2023-mf}. After every few cases, autoethnographers are given a chance to contribute, cull, and cluster the post-it notes that represent features from cases. This pause-point serves as a member-check among autoethnographers. Once this analysis is complete, researchers are encouraged to circulate any subsequent materials---such as a publication manuscript---with autoethnographers as a final member-check.

\subsection{Member Checked ARCs}
Asynchronous Remote Community methods (ARCs) were introduced by~\citet{MacLeod2017-xw} to engage with distributed and minority populations, such as patients with rare diseases, stigmatized conditions~\cite{Maestre2018-bw}, and marginalized identities~\cite{DeVito2021-yv}. ARCs are a powerful tool at the intersection of participatory design~\cite{Sanders2010-ha} and action-research~\cite{Avison1999-ov} because they overcome the access barriers of face-to-face group-based activities~\cite{Prabhakar2017-lt}. Moreover, ARCs can provide a new level of safety by leveraging online pseudonyms, burner profiles, and private groups~\citet{Maestre2018-bw}.~\citet{DeVito2021-yv} were able to create a safe discussion space with anticipatory standard-setting and on-the-ground moderation in their ARC deployment. In this section, we leverage the promising features of ARCs to afford epistemic autonomy to research participants. Specifically, we call for a new variant of ARCs, \textbf{member checked ARCs}, that hold participant knowledge as a sacred outcome rather than data to be extracted. 

\begin{figure}
    \centering
    \includegraphics[width=0.9\linewidth]{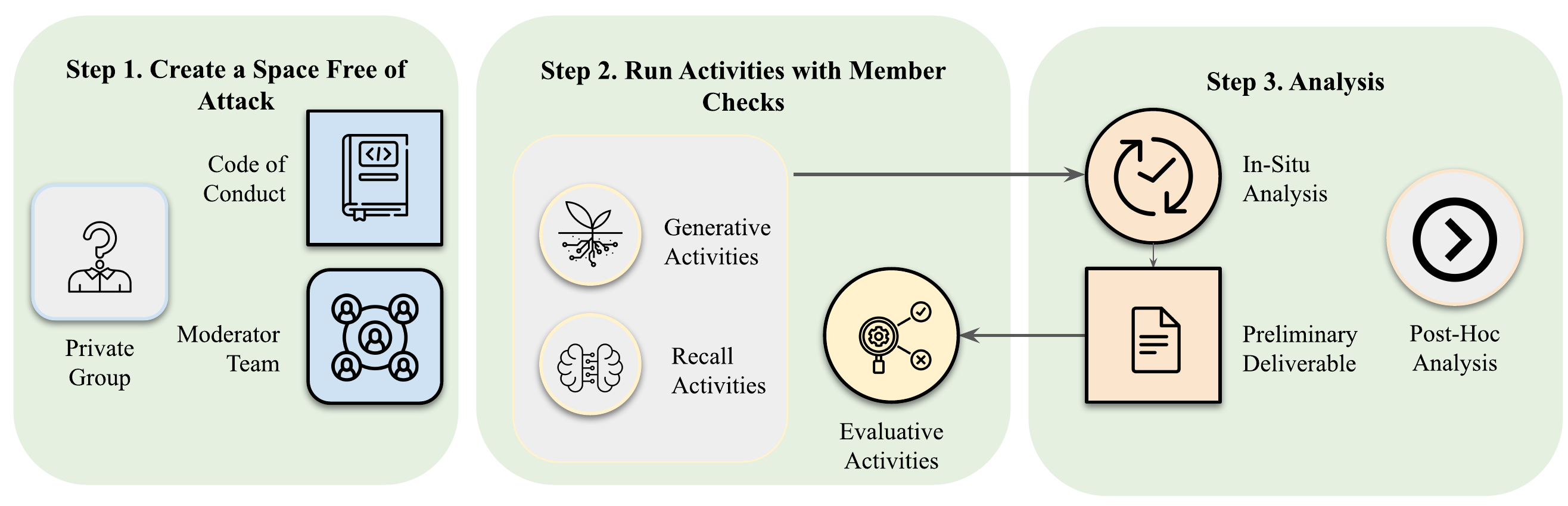}
    \Description{Three-stage diagram of the steps to implement a member-checked ARC}
    \caption{Steps and associated documents for member checked ARCs. Building on the promise of ARCs to provide private spaces, various activities, and rich analysis (in grey), member-checked ARCs involve a cyclical process of member-checking preliminary analyses (in color).}
    \label{fig:member-arc}
\end{figure}

Typically used in methods that have a large amount of researcher interpretation, such as grounded theory~\cite{Braun2006-eg}, member checking is the process of returning findings to research participants in order to check for accuracy and resonance~\cite{Nowell2017-cx}. However, member checking has its limitations in fields such as HCI, where research papers are written with a certain jargon and scope. For example,~\citet{Motulsky2021-ce} note how simply circulating a final manuscript to non-expert research participants is a limited method of member checking. To that end, we define member checking as the iterative process of returning researcher interpretation to participants---what~\citet{Doyle2007-fh} describes as the hermeneutic cycle of research. Below, we translate~\citet{MacLeod2017-xw}'s steps of designing an ARC to include consistent member checking (see Figure~\ref{fig:member-arc}). 

\edit{Member checked ARCs afford epistemic autonomy is a few ways. First, similar to traditional ARCs, member checked ARCs afford testimonial autonomy through safety; the first step is for researchers to create spaces where testimony can be given free of attack by using privacy features, standard setting, and community moderation.} Hermeneutically, member-checked ARCs require that researchers are consistently articulating their interpretations to the research participants and adapting from there. Recall that community research is a translational effort and hermeneutical autonomy is the act of starting from a vocabulary \textit{of the community}. Member-checked ARCs allow researchers to balance this translational effort while ensuring community resonance.

\subsubsection{Step 1. Create a Space Free of Attack} Asynchronous Remote Community methods start by having an online space where the community can engage in planned activities. As prior work indicates, online spaces are a double-edged sword for marginalized communities; they simultaneously provide collective visible spaces for building community and visible spaces for potential harassment~\cite{Scheuerman2018-uo, DeVito2022-ee, Ajmani2024-ll}. \edit{Recall that testimonial autonomy requires spaces free from attack. Therefore, ARCs that afford testimonial autonomy must be \textit{safe spaces by design}.} Inspired by previous ARCs with the LGBTQ$+$ and trans communities~\cite{walker_more_2020}, safety requires the researcher to create three artifacts. First, the private group (e.g., Facebook group, discord server, etc.) must have affordances for anonymity, such as the ability to create accounts with no connection to personal profiles, such as Finstas~\cite{Huang2022-az, Taber2020-or}. Second, it is on the researcher to create a code of conduct for the group. It is important to remember that communities are still vulnerable to internal stigma. For example, the LGBTQ$+$ community often perpetrates transmisogyny and trans erasure~\cite{Weiss2012-va}. Moreover, intersectionality~\cite{Collins2020-bx} notes that there are axes of marginalization and power that play out within communities. For example, women of color are often not afforded the same credibility as other women~\cite{Taylor2017-vm, Gray2021-wd, Collins2017-mt}. Therefore, even ARCs that bring together similar identities need proactive standard setting to create a safe, stigma-free space. Finally, while standard setting is a proactive mechanism for creating a safe space, researchers should still plan for standard violations. To that end, member-checked ARCs must have a moderator team that maintains and enforces standards. Ideally, moderators are of the community and afforded due credibility by the researcher. Here, we call on recent work that outlines moderation strategies based on care and power~\cite{Gilbert2023-pg}. This intersectional model of moderation accounts for the inherent power imbalance among participants, researchers, and the various identities ARCs bring together.

\subsubsection{Step 2. Run Activities with Member Checks}
Formative work on ARCs has studied the variety of feasible asynchronous design activities~\cite{walker_more_2020, DeVito2021-yv, Maestre2018-bw}. \citet{MacLeod2017-xw} describe activities as either recall-based or generative. For example, participants were asked to rank a list of problems by recalling how much each affected them personally. Later in the ARC, they were asked to generate solutions to the highest ranked problems. In member-checked ARCs we propose another activity type: evaluative activities. In evaluative activities, ARC members are asked to evaluate a deliverable the researcher has made based on preliminary analysis. In other words, the researcher articulates their interpretation through a document, prototype, or diagram and checks it with ARC members.~\citet{DeVito2024-vd} explored similar activities through ``reconciliation rounds'' where big researcher decisions and tradeoffs were presented back to participants. In this case, member activities and researcher analysis are not asynchronous procedures. Rather, researchers are required to do preliminary analyses during the ARC, create a document that is interpretable to ARC members, and solicit feedback. This cyclical process of researcher analysis, deliverable creation, and member evaluation is crucial to affording epistemic autonomy.
\subsubsection{Step 3. Analysis}
In their original form, ARCs happen over weeks or months. Once the ARC has concluded, the researchers use multiple techniques to generate results. For example,~\citet{Maestre2018-bw} quantified participant engagement along their eight-week ARC. Additionally, they used qualitative open coding and triangulation techniques to measure participant feedback and consistency. In member-checked ARCs, these analysis techniques are post-hoc insofar as they happen after the ARC is complete. These analyses lead to rich and holistic findings about the ARC as a multi-week method.

However, relying solely on post-hoc analyses does not leave room for members to check a researcher's interpretation. As~\citet{Doyle2007-fh} articulates, member checking is a method for \textit{active} participation in the research process. Recall that a researcher's interpretation is a powerful tool in affording or violating the epistemic autonomy of participants. Therefore, it is crucial that researchers couple this post-hoc analysis with lighter-weight preliminary analyses during the ARC that lead to artifacts. For example, researcher memos~\cite{Lempert2007-ms}, mind maps~\cite{Crowe2012-tl}, or initial write-ups are valuable parts of the analysis that ARC participants can evaluate. With these initial deliverables, we hope to participant-researcher dynamic with consistent member-checking.

\section{Discussion}
Without exactly calling it as such, researchers
in our field already think about epistemic autonomy and have successfully generated change. User-centered design, one of the first attempts to center the needs of the people we research and design for~\cite{Norman1986-eb}, has spurred the entire discipline of User Experience (UX) work in both research and industry. From there, researchers asked: \textit{can we give users more say?} and created the participatory design toolkit~\cite{Sanders2002-tn}. Specifically at CHI, researchers asked: \textit{can we bring participants closer to the design process?} and evolved participatory design into co-design~\cite{Vines2013-gj}. In this paper, we once again ask: \textit{can we empower participants more?} Specifically, we focus on giving due authority to those who were left in the margins with previous attempts to center the user. We mainly highlight the transfeminine experience because our concept of \textit{autonomy} belongs to them. However, we believe that affording epistemic autonomy to research participants will systemically raise discounted voices. 

CHI is uniquely poised to bake in the values of epistemic autonomy with groups with intersecting, marginalized identities. CHI has a rich history of studying the intersections of gender, technology, and research. Over a decade ago,~\citet{Bardzell2011-ln} proposed principles of Feminist HCI. Since then, researchers have used this lens of gender and power to unpack women's health technology~\cite{D-Ignazio2016-yx, Sondergaard2023-lg, O-Neill2024-ra}, dating apps~\cite{Zytko2023-ml, Aljasim2023-lr, DeVito2024-vd}, and even seemingly gender-neutral platforms, like Wikipedia~\cite{Menking2021-ic, Menking2019-cz}. CHI has already committed to understanding how the politics of identity affect both user experiences~\cite{DeVito2022-ee, DeVito2018-gr} and researcher experiences~\cite{Erete2021-ul, Ogbonnaya-Ogburu2020-wt}. We propose CHI strengthen this commitment by giving transfems, women of color, and other marginalized communities the authority to govern how they are known. In this paper, we propose variants of two common methods in CHI, autoethnography~\cite{Rapp2018-im, Fassl2023-ur, OKane2014-fa, Lucero2018-gl, Ajmani2024-ll} and asynchronous remote communities~\cite{Maestre2018-bw, walker_more_2020, DeVito2021-yv} to demonstrate how close CHI already is to further center the user's voice. Moreover, we provide an account of a CHI researcher attempting to afford epistemic autonomy in her own research. We hope this vignette serves as a foundation for inspiration and future discussion of research practices. While epistemic autonomy is a fundamental restructuring of our core beliefs, it is not a complete restructuring of our research practices. In fact, our practices are a robust foundation. Therefore, the burden to lead the charge lies on CHI.

Finally, we recognize that HCI is a field of many paradigms that have all made HCI what it is today. For example, constructivist and interpretivist paradigms inspired the participatory design methodology~\cite{Sanders2010-ha}. Meanwhile, positivist paradigms focused on valid measurement have given us valuable measurement tools for determining application usability, fairness, and model bias~\cite{Wobbrock2016-sx}. Critical lenses have been applied to multiple paradigms to interrogate how societal power dynamics affect our current research practices~\cite{Ajmani2023-cp, Pierre2021-zz}. We believe this multi-dimensional approach to paradigms in HCI gives the field strength to serve as what Blackwell calls an ``inter-discipline~\cite{Blackwell2015-ei}.'' And yet, calls for justice and feminist-oriented research continue to go unanswered in our field~\cite{Chivukula2020-gk, Chordia2024-ff}. Participation continues to be a debated term as research often breeds superficial involvement of research participants rather than meaningful ownership~\cite{Pierre2021-zz, Delgado2021-ec}. In this paper, we outline epistemic autonomy as a paradigm that specifically focuses the epistemic justice of research participants, particularly those with intersecting marginalized identities. While epistemic autonomy may not be appropriate for all HCI research, the fundamental premise holds as a commitment CHI must make. To achieve justice in our research practices, we must allow research participants to govern the knowledge they share with us. To that end, we believe epistemic autonomy belongs in every CHI researcher's toolkit.

\edit{Beyond CHI, our work is in concert with the other fields that question current power structures. For example, a researcher-participant relationship rooted in epistemic autonomy is a progression of Black feminist articulations of relationality~\cite{Collins2020-bx, Collins2017-mt}. Furthermore, the idea of moving beyond research participant inclusion and towards ideals of epistemic justice is central to disability activism~\cite{Ymous2020-uc}. In this paper, we embrace the sentiment from disability activists \textit{``nothing about us without us.''} As CHI translates the calls from these critical disciplines into practice~\cite{Bardzell2011-ln, Ogbonnaya-Ogburu2020-wt, Schlesinger2017-af}, we must also turn our gaze inwards. We hope that epistemic autonomy serves as a bridge between critical scholarship and CHI researchers.}

In conclusion, epistemic autonomy as a research paradigm builds on the work CHI is already doing and calibrates us toward where we want to go. Notably, incorporating epistemic autonomy into research practices often requires slight variations to popular HCI methods, such as autoethnography and ARCs. We encourage future work to explore other popular methods and analysis techniques as foundations for furthering epistemic autonomy. This methodological work is crucial to empowering communities that we have often unconsciously shoved aside, such as the transfem community. 

\section{Limitations and Future Work}
At its core, this work is a call for change in HCI. Specifically, we call to further center participant knowledge through our research beliefs and practices. A main limitation of any call for change is that change requires effort. More often than not, this effort falls on those already in the margins of our research field~\cite{Erete2021-ul}. We recognize that a natural interpretation of this work is for the trans community to do more work in HCI. We caution against this interpretation and suggest movements towards member research in HCI~\cite{Adler1987-ju}. Future work could explore how to afford epistemic autonomy and train transgender women, for example, to be research contributors.

Moreover, we focus on the transfeminine experience in this paper. Autonomy---particularly in terms of knowledge---has been repeatedly articulated, iterated upon, and codified in transfeminist philosophy and activism~\cite{Koyama2020-zi, Serano2007-yu}. Trans women are a paramount case of intersecting axes of marginalized. As described in our stories, even within trans-oriented research trans women's perspectives get unfairly discounted. The two main philosophies we pull from, transfeminism and epistemic injustice, have been proven to be in solidarity with other marginalized communities~\cite{Krell2017-iv, Collins2017-mt, Heyes2003-ip}. We encourage future work to empirically test this generalizability. For example, inspired by~\citet{Ajmani2024-ll}'s collaborative autoethnography, future work could use scaffolded autoethnography to explore indigenous experiences in online communities. We encourage future work to broaden the scope and applicability of epistemic autonomy.




\bibliographystyle{ACM-Reference-Format}
\bibliography{main}


\end{document}